\documentclass[finallayout,an,fleqn]{w-art}
\usepackage{times}
\usepackage{w-thm}
\theoremstyle{plain}

\theoremstyle{definition}

\usepackage[]{graphicx}
\chardef\bslash=`\\ 

\catcode`\@=11 
\def\@versim#1#2{\vcenter{\offinterlineskip
        \ialign{$\m@th#1\hfil##\hfil$\crcr#2\crcr\sim\crcr } }}
\def\lsim{\mathrel{\mathpalette\@versim<}}
\def\gsim{\mathrel{\mathpalette\@versim>}}
\def\mpy{M_\odot \ {\rm yr^{-1}}}
\def\rad{\rm \ rad \ m^{-2}}
\begin{document}

\DOIsuffix{theDOIsuffix}
\Volume{324}
\Issue{S1}
\Copyrightissue{S1}
\Month{01}
\Year{2003}
\pagespan{3}{}
\Receiveddate{15 November 2002}
\Reviseddate{30 November 2002}
\Accepteddate{2 December 2002}
\Dateposted{3 December 2002}



\title[Accretion Models for Sgr A*]{Radiatively Inefficient Accretion Flow Models of Sgr A*}


\author[Quataert]{Eliot Quataert\footnote{e-mail: {\sf eliot@astron.berkeley.edu}; Phone: 510-642-3792}\inst{1}} \address[\inst{1}]{Astronomy Dept., 601 Campbell Hall, UC Berkeley, Berkeley, CA 94720}
\begin{abstract}

I review radiatively inefficient accretion flow models for the
$\approx 2.6 \times 10^6 M_\odot$ black hole (BH) in the Galactic
Center.  I argue for a 'concordance model' of Sgr A*: both theory and
observations suggest that hot ambient gas around the BH is accreted at
a rate $\sim 10^{-8} \mpy$, much less than the canonical Bondi rate.
I interpret {\it Chandra} observations of Sgr A* in the context of
such a model: (1) the extended 'quiescent' X-ray emission is due to
thermal bremsstrahlung from gas in the vicinity of the Bondi accretion
radius, and (2) the $\sim 10^4$ second long X-ray flares are due to
synchrotron or Inverse-Compton emission by non-thermal electrons
accelerated in the inner $\sim 10$ Schwarzschild radii of the
accretion flow.
\end{abstract}

\maketitle                   







\section{Introduction}

The case for a $\approx 2.6 \times 10^6 M_\odot$ black hole (BH)
coincident with the radio source Sagittarius A* in the Galactic Center
(GC) is now compelling (e.g., Sch\"odel et al. 2002; Ghez et
al. 2003).  This only emphasizes the long-standing puzzle that the
luminosity from the GC is remarkably low given the presence of a
massive black hole. The resolution of this puzzle must lie in how gas
from the ambient medium accretes onto the central BH.  In these
proceedings I review accretion models and their application to Sgr A*.




A unique feature of the Galactic Center is our ability to constrain
the dynamics of gas quite close to the black hole (relative to other
systems), thus providing additional boundary conditions on, and much
less freedom for, theoretical models.  A canonical formulation of
these constraints is the Bondi accretion estimate for the rate at
which the BH gravitationally captures surrounding gas (Bondi 1952; see
Melia 1992 for an early application to Sgr A*). Given relatively
uniformly distributed matter with an ambient density $\rho_0$ and an
ambient sound speed $c_0$, the sphere of influence of a BH of mass $M$
extends out to $R_{acc} \approx GM/c^2_0$.  The accretion rate of this
gas onto the central BH, in the absence of angular momentum and
magnetic fields, is then $\dot M_B \approx \pi R_{acc}^2 \rho_0 c_0$.

{\it Chandra} observations of the GC detect extended diffuse emission
within $1-10''$ of the BH (Baganoff et al. 2003a).  This emission
likely arises from hot gas produced when the stellar winds from
massive stars in the GC collide and shock (e.g., the He I cluster;
Krabbe et al. 1991).  Interpreted as such, the inferred gas density
and temperature are $\approx 20$ cm$^{-3}$ and $\approx 1.3$ keV on
$10''$ scales, and $\approx 100$ cm$^{-3}$ and $\approx 2$ keV on
$\approx 1''$ scales (see also Fig. \ref{fig:3}).  The corresponding
Bondi accretion radius is $R_{acc} \approx 0.04 {\rm pc} \approx 1''$
and the Bondi accretion rate is $\dot M_B \approx 10^{-5} M_\odot$
yr$^{-1}$.\footnote{This is much less than the total mass loss rate
from stars in the GC ($\approx 10^{-3} \mpy$; Najarro et al. 1997),
implying that there should also be a global outflow of hot gas from
the central parsec (a GC 'wind').}  If gas were accreted at this rate
onto the BH via a geometrically thin, optically thick accretion disk
(Shakura \& Sunyaev 1973), a model that has been extensively and
successfully applied to {luminous} accreting sources (e.g., Kortakar
\& Blaes 1999), the expected luminosity would be $L \approx 0.1 \dot
M_B c^2 \sim 10^{41}$ ergs s$^{-1}$, larger than the observed
luminosity by a factor of $\sim 10^5$.  This is the strongest argument
against a thin disk in Sgr A*.  An additional argument is the absence
of any disk-like blackbody emission component in the spectrum of Sgr
A*.  If the putative disk were to extend all the way down to the BH,
its accretion rate would have to be $\lsim 10^{-10} M_\odot$ yr$^{-1}
\approx 10^{-5} \dot M_B$ to satisfy infrared limits (e.g., Narayan
2002; see his Fig. 2).

One possible caveat to the Bondi analysis is that there is far more
(by mass) cold molecular gas than hot X-ray emitting gas in the
central $1-10$ parsecs of the GC (e.g., Herrnstein \& Ho 2002).  It is
unclear how close to the BH the molecular gas extends and whether it
is important for the dynamics of gas accreting onto the BH.  In what
follows I ignore this component, but see Nayakshin (2003) for a
different view.

As emphasized above, the inferred low efficiency of Sgr A* is the
strongest argument against accretion proceeding via a thin accretion
disk.  Instead, the observations favor models in which very little of
the gravitational potential energy of the inflowing gas is radiated
away.
I will refer to such models as radiatively inefficient accretion flows
(RIAFs).  In the next section (\S2) I summarize the properties of
RIAFs.  I then apply these models to the GC (\S3), emphasizing the
interpretation of radio and X-ray observations of Sgr A*.  Finally, I
conclude with a brief summary (\S4).

\section{Radiatively Inefficient Accretion Flows}

RIAFs describe the dynamics of rotating accretion flows in which $L
\ll 0.1 \dot M c^2$, i.e., very little energy generated by accretion
is radiated away (e.g., Ichimaru 1977; Rees et al. 1982; Narayan \& Yi
1994).  Instead, the gravitational potential energy released by
turbulent stresses in the accretion flow is stored as thermal energy.
As a result, the accreting gas is very hot, with a characteristic
thermal energy comparable to its gravitational potential energy; close
to the BH this implies $T \sim GMm_p/3kR \sim 0.1 m_p c^2/k \sim
10^{12}$ K.  At such temperatures, and for gas densities appropriate
to systems like the GC, the Coulomb collision time is much longer than
the time it takes gas to flow into the BH.  The accretion flow then
develops a two-temperature structure with the protons likely hotter
than the electrons: $T_p \sim 10^{12}$ K $\gsim T_e \sim
10^{10}-10^{12}$ K.  The precise electron temperature is uncertain but
important since electrons produce the radiation that we see.  The
electron temperature depends on how and to what extent they are heated
by processes such as shocks, MHD turbulence, and reconnection (see,
e.g., Quataert \& Gruzinov 1999).  Note that because collisions are
unimportant one would not expect the electron distribution function to
be thermal.


Advection-Dominated Accretion Flows (ADAFs) are a simple analytical
model for the dynamics of RIAFs; they predict that the structure of
the flow is in some ways similar to spherical Bondi accretion, despite
the fact that angular momentum and viscosity are important (e.g.,
Ichimaru 1977; Narayan \& Yi 1994).  In ADAF models the gas rotates at
$\Omega \approx 0.3-0.5 \ \Omega_K$, where $\Omega_K = \sqrt{GM/R^3}$
is the Keplerian rotation rate.  Because the flow is hot, pressure
forces are also important and the inflowing gas is geometrically quite
``thick,'' with a scale height $H \approx R$ at every radius.  The
radial velocity in the flow is given by $v_R \approx \alpha c_s
\approx \alpha v_K$ where $\alpha$ is the dimensionless viscosity
parameter, $c_s$ is the sound speed in the flow and $v_K = R \Omega_K
\propto R^{-1/2}$.  Conservation of mass on spherical shells then
implies that the density scales as $\rho \propto R^{-3/2}$, the
characteristic scaling for spherical accretion.  {\it ADAF models also
predict that, even in the presence of rotation, the rate at which gas
accretes onto the BH from an ambient medium is comparable to the Bondi
accretion rate: $\dot M_{ADAF} \sim \dot M_B$}.\footnote{A more
accurate estimate might be $\dot M_{ADAF} \approx \alpha \dot M_B$
(e.g., Narayan 2002).  The factor of $\alpha$ arises because the
inflow velocity of gas at the Bondi accretion radius is $\approx
\alpha c_s$ in ADAF models while it is $\approx c_s$ in Bondi models.
Thus for a fixed density in the ambient medium, the accretion rate in
an ADAF will be smaller by a factor of $\alpha$.}  {\it Thus in ADAF
models the low luminosity of Sgr A* is due to a very low radiative
efficiency $\sim 10^{-6}$.}

With the advent of global, time-dependent, numerical simulations of
accretion flows, it has become possible to numerically simulate RIAFs
and test the ADAF predictions.  Note that RIAFs are, in one sense, the
easiest flows to simulate since (1) no treatment of radiation or
radiative transfer is needed and (2) the flow is ``thick'' with $H
\sim R$, so there is no difficult-to-simulate separation of scales
like in a thin disk.  On the other hand, for a system like the GC, the
proton-electron collision time close to the BH is $\sim 6$ orders of
magnitude longer than the inflow time of the gas.  Thus the fluid
approximation used by all simulations to date is suspect (a kinetic
treatment should be used; see Quataert et al. 2002).  Whether this
introduces qualitative or merely quantitative errors in the results is
unknown.  I suspect the latter.

The key result from nearly all simulations to date is that {\it $\dot
M \ll \dot M_{B}$, i.e., very little mass available at large radii
actually accretes onto the black hole} (e.g., Stone, Pringle, \&
Begelman 1999; Igumenshchev \& Abramowicz 1999; 2000; Igumenshchev et
al. 2000; Stone \& Pringle 2001; Hawley \& Balbus 2002; Igumenshchev
et al. 2003).  Another way to state this result is that the radial
density profile in the flow is much ``flatter'' than the ADAF
predictions: for a given gas density at large distances from the black
hole (e.g., measured by {\it Chandra} on 1'' scales in the GC), the
density close to the BH is much {\it less} than the ADAF or Bondi
predictions.  Following a proposal due to Blandford \& Begelman
(1999), we can parameterize the density profile of RIAFs with a
parameter $p$, where $\rho \propto R^{-3/2 + p}$.  With this
parameterization, the rate at which gas is actually accreted into the
BH is $\sim (R_{in}/R_{out})^p \dot M_B$, where $R_{in} \sim R_S$ is
the inner radius of the flow and $R_{out} \sim R_{acc}$ is the outer
radius.  Values of $p \approx 1/2-1$, rather than $p = 0$ predicted by
ADAF models, are favored by the simulations.  {\it This implies that,
within the context of RIAF models, a low accretion rate, $\dot M \ll
\dot M_B$, rather than just a low efficiency, is a major factor in the
faintness of Sgr A*.}


Since the focus of this review is the {application} of RIAF models to
Sgr A*, I will not dwell on why the simulations differ so
significantly from the ADAF predictions.  A brief discussion is,
however, in order.  The following material will not be used in later
sections so uninterested readers can move directly to \S3.

In RIAFs, the inflowing gas is heated at a rate $\approx 0.1 \dot M
c^2$, as required by the release of gravitational potential energy in
a differentially rotating accretion flow.  In ADAF models this energy
is stored as thermal energy and carried into the BH.  As noted above,
the inflowing gas is then very hot with a sound speed at any radius
comparable to the escape speed from the BH's potential well.  ADAF
models are therefore prone to developing outflows (see Narayan \& Yi
1994 or Blandford \& Begelman 1999 for a more formal discussion).
This led Blandford \& Begelman (1999) to propose that, in the absence
of radiation, the gravitational binding energy of the accreted gas
must be lost through some other (non-radiative) means.  Otherwise the
inflowing gas is not sufficiently bound to the BH to accrete.  The
numerical simulations to date are broadly consistent with this
hypothesis; e.g., the gas temperature in the simulations is generally
a factor of few-5 less than in ADAF models implying that the gas is
indeed more strongly bound to the BH.

The non-radiative energy loss can take one of two forms: (1) efficient
turbulent transport of energy through the accretion flow to large
radii or (2) a global outflow ('wind') that carries away the binding
energy of the accreted matter.  In hydrodynamic simulations (e.g.,
Stone et al. 1999; Igumenshchev \& Abramowicz 1999) or in MHD
simulations with relatively 'weak' magnetic fields ($\beta \gsim
10-100$, where $\beta$ is the ratio of the gas pressure to the
magnetic pressure), convective transport of energy and angular
momentum dominates the dynamics of the accretion flow (Narayan et
al. 2002; Igumenshchev et al. 2003; see Quataert \& Gruzinov 2000a and
Narayan et al. 2000 for such 'convection-dominated accretion flow'
models).  The convective luminosity through a spherical shell at
radius $R$ is $\propto \rho v_c^3 R^2$ where $v_c$ is the turbulent
convective velocity.  Since the flow is hot, $v_c \propto c_s \propto
v_K \propto R^{-1/2}$. A constant flow of gravitational binding energy
from small to large radii therefore requires $\rho \propto R^{-1/2}$;
i.e., $p = 1$ instead of $p = 0$ as in ADAF models.\footnote{The
convective energy flux could launch a thermally driven wind from large
radii $\sim R_{acc}$ or from the surface layers of the accretion flow.
Thus there is not necessarily a clean distinction between global
energy transport by turbulence and mass outflow as mechanisms for
'non-radiative' energy loss.  Both processes are related and, in
particular, the former can drive the latter.}

In contrast to the hydrodynamic results, in MHD simulations with
strong magnetic fields ($\beta \lsim 10$), MHD turbulence dominates
the flow dynamics and convection is unimportant (e.g., Stone \&
Pringle 2001; Hawley \& Balbus 2002; Igumenshchev et al. 2003).  Stone
\& Pringle (2001) and Hawley \& Balbus (2002) find that most of the
inflowing gas is lost to a magnetically driven wind.  Igumenshchev et
al. (2003), on the other hand, find a much more complex flow
configuration, though still with a very small accretion rate.

\section{RIAF Models Applied to Sgr A*}

\begin{figure}[htb]
\begin{minipage}[t]{.45\textwidth}
\includegraphics[width=\textwidth]{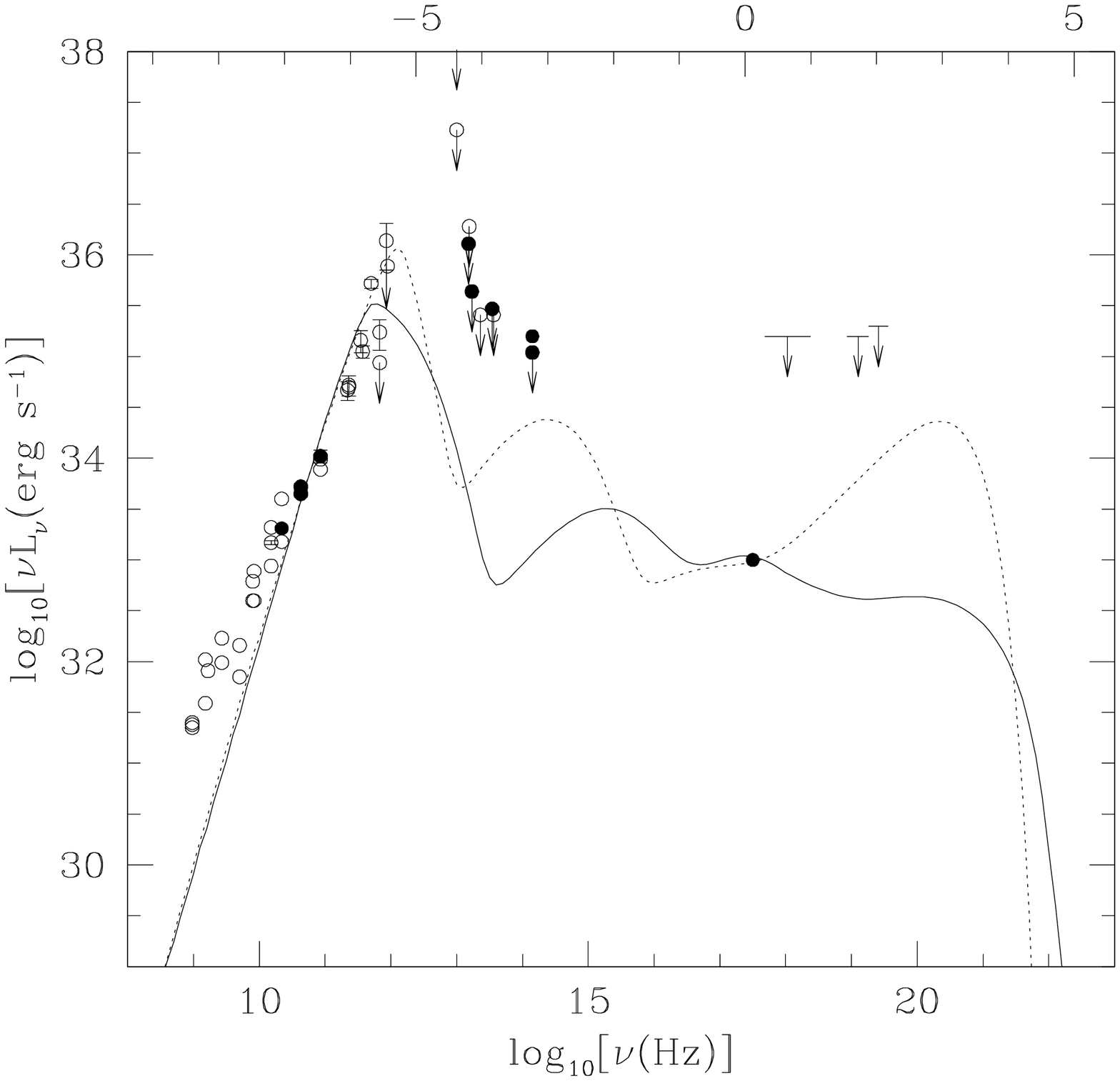}
\caption{Two 'baseline' RIAF models of Sgr A* that reproduce the
quiescent {\it Chandra} flux (the {\it Chandra} spectrum is discussed
in Figs \ref{fig:2} and \ref{fig:3}).  {\it Dotted line}: a RIAF model
with $p = 0$, $\dot M \approx \dot M_B$ and $\delta = 0.01$ ($\delta
\equiv$ fraction of turbulent energy heating electrons). This is an
ADAF-type model.  {\it Solid line}: a RIAF with $p = 0.5$ and a net
accretion rate into the BH of $\dot M \approx 10^{-8} \mpy \ll \dot
M_B$; $\delta = 0.5$.  This Figure is based on Quataert \& Narayan
(1999).}

\label{fig:1}
\end{minipage}
\hfil
\begin{minipage}[t]{.45\textwidth}
\vspace{-6cm}
\includegraphics[angle = -90, width=\textwidth]{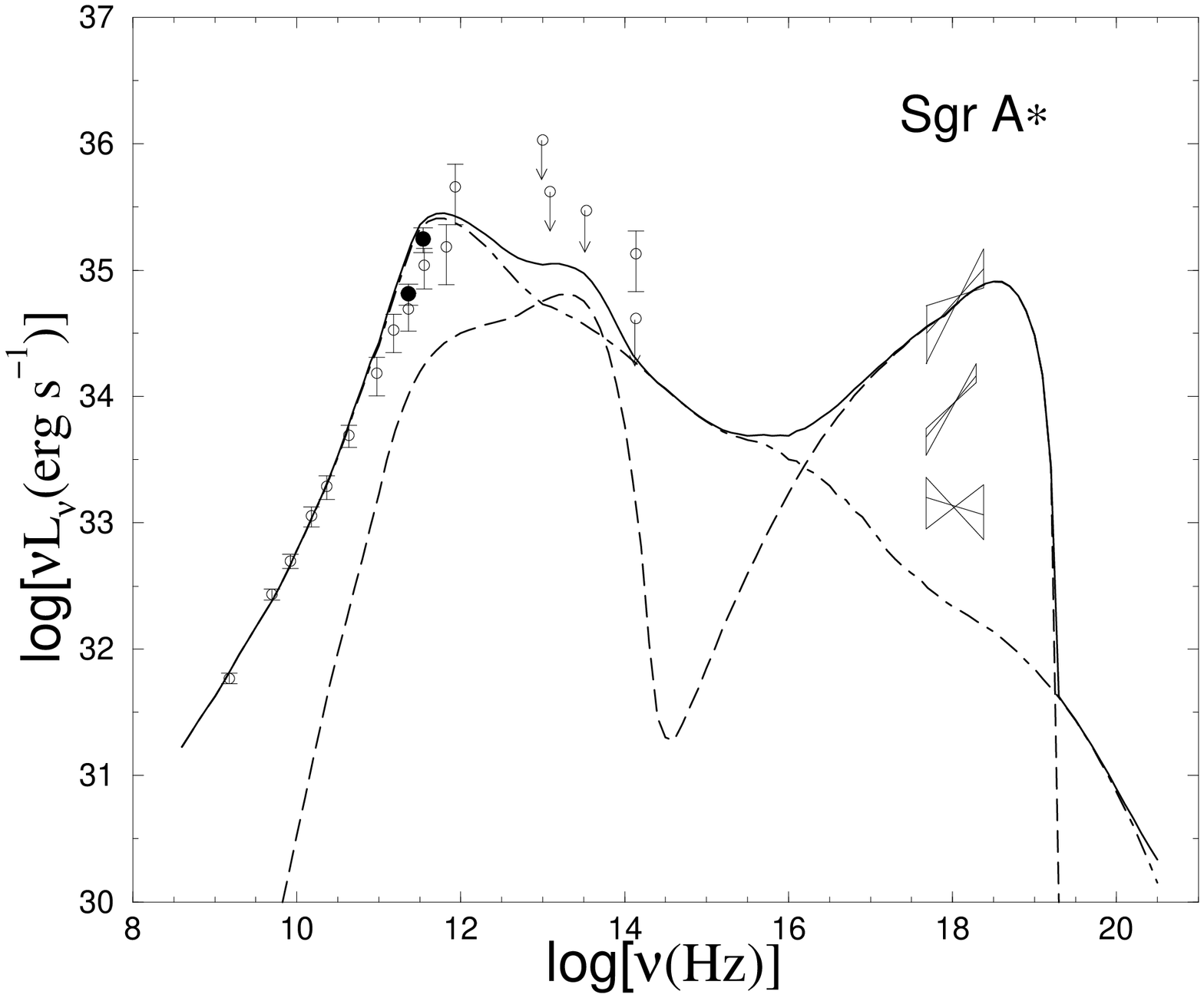}
\vspace{0.17cm}
\caption{{\it Dot-dashed}: A RIAF with $p \approx 0.4$ and $\delta
\approx 0.5$.  There is also a power law distribution of electrons
with $n(\gamma) \propto \gamma^{-3.5}$ and $\approx 2 \%$ of the
electron thermal energy.  Power-law electrons produce the low
freq. radio emission not accounted for in Fig. \ref{fig:1}.  {\it
Dashed line}: Inverse-Compton model for the {\it Chandra} X-ray flare
(see text for details).  {\it Solid line:} Total emission.  X-ray
error bars are (from top to bottom): Oct. 2000 flare, average flare,
\& quiescent emission.  This Figure is based on Yuan et al. (2003).}
\label{fig:2}
\end{minipage}
\end{figure}

A number of authors have used RIAF models, and in particular ADAF
models, to explain the observed properties of Sgr A* (e.g., Narayan et
al. 1995, 1998; Manmoto et al. 1997).  These calculations have shown
that an ADAF model accreting at the observationally inferred rate,
$\dot M \sim \alpha \dot M_B$, can roughly account for the observed
luminosity and spectrum of Sgr A*.  The key constraint is that the
fraction of the turbulent energy that heats the electrons, $\equiv
\delta$, must be small $\lsim 0.01$, so as to not overproduce the
observed luminosity.  Equivalently the electron temperature close to
the BH must be $\lsim 10^{10}$ K $\ll T_p \approx 10^{12}$ K.  An
example of such a model in shown by the dotted line in Figure
\ref{fig:1}.  The model roughly reproduces the observed sub-mm
emission, satisfies the IR limits, and produces an X-ray luminosity
comparable to that seen by {\it Chandra} in the quiescent
(non-flaring) state (I discuss the {\it Chandra} observations in more
detail below).  It does, however, significantly underproduce the lower
frequency radio emission.  Since the lower frequency radio emission in
Sgr A* is phenomenologically similar to that of other AGN, a natural
interpretation is that a jet is present and produces the radio
emission (e.g., Falcke \& Markoff 2000, Yuan et al. 2001, and
references therein).  Alternatively, the results in Figure \ref{fig:1}
assume purely thermal electrons, for which there is no good
justification.  As Figure \ref{fig:2} shows, even a small population
of nonthermal electrons in the accretion flow can produce the radio
emission (e.g., Mahadevan 1998; Ozel et al. 2000).

Quataert \& Narayan (1999) showed that a much broader class of RIAF
models could also account for the observed properties of Sgr A*.
Specifically, the low accretion rate models favored theoretically
(\S2) can reproduce the observations as well.  The key requirement is
that the electrons must be hotter so as to produce more emission even
though $\dot M$ and the gas density are lower. An example of such a
model is shown by the solid line in Figure \ref{fig:1}: $p = 0.5$ from
$R_{out} = 10^5 R_S$ down to $R_{in} = R_S$, implying that the
accretion rate into the BH is much smaller than the Bondi rate ($\sim
10^{-8} \mpy$).  The electron temperature close to the BH is $\sim
10^{11}$ K, rather than $\sim 10^{10}$ K as in ADAF models.

To first approximation, the two models in Figure \ref{fig:1} reproduce
the observed spectrum of Sgr A* equally ``well'' (or poorly, depending
on one's vantage point).  {\it However, Aitken et al. (2000) and Bower
et al.'s (2003) detection of $\approx 10 \%$ linear polarization in
the sub-mm (230 GHz) emission from Sgr A* argues strongly for low
accretion rate models with $\dot M \lsim 10^{-8} \mpy$.}


Beckert \& Falcke (2002, 2003; see also Ruszkowski \& Begelman 2002)
present detailed models for the radio polarization of Sgr A* (both
linear and circular).  Here I summarize the constraints imposed by the
linear polarization detection: accretion at $\sim \dot M_B$ implies a
much higher gas density and magnetic field strength close to the black
hole than accretion at $\ll \dot M_B$.  Faraday rotation is therefore
much stronger (e.g., Quataert \& Gruzinov 2000b; Agol 2000).  In
models with $\dot M \sim \dot M_B$, the rotation measure is $\gsim
10^{10} \rad$ in the region of the flow ($\lsim 10-100 R_S$) where the
sub-mm emission is produced.  This leads to a Faraday rotation angle
$\sim 10^5$ {\rm rad} at $\sim 100$ GHz.  This large rotation angle
implies that intrinsically linearly polarized synchrotron emission
would be depolarized propagating to the observer over most of the
radio to infrared spectrum.  By contrast, Bower et al. (2003) find
that $RM \lsim 10^6 \rad$ in their linear polarization detection at
$230$ GHz.  Models with $\dot M \ll \dot M_B$ can satisfy this
constraint because the density and magnetic field strength close to
the BH are much smaller. For example, for $\dot M \approx 10^{-8}
\mpy$, the net rotation measure through the accretion flow is $RM
\approx 10^6 \rad$, consistent with the observational constraint.
Thus the observed detection of linear polarization in the sub-mm
emission of Sgr A* argues for a low accretion rate $\lsim 10^{-8} \mpy
\ll \dot M_B$.  One way out of this conclusion is to posit that the
magnetic field undergoes so many reversals along the line of sight that
the net Faraday rotation is $\gsim 10^5$ times smaller than these
simple estimates (Ruszkowski \& Begelman 2002).  I regard this as very
unlikely, but am not aware of a direct observational argument against
this possibility at the present time.



In light of the above arguments the rest of my discussion centers on
models with $\dot M \ll \dot M_B$: these 'concordance' models are both
theoretically favored and satisfy the rotation measure constraint from
the linear polarization detection.  I focus on interpreting the {\it
Chandra} X-ray observations of Sgr A*.  Much of this material is based
on Quataert (2002) and Yuan, Quataert, \& Narayan (2003).

{\it Chandra} observations reveal that there are two components to the
X-ray emission coincident with Sgr A* (Baganoff et al. 2001, 2003ab):
(1) a 'baseline' X-ray flux with a nearly constant luminosity $\approx
2 \times 10^{33}$ ergs s$^{-1}$ and a soft spectrum (photon index
$\Gamma \approx 2.7$, where $\nu L_\nu \propto \nu^{2 - \Gamma}$).
This component is clearly extended with a size of $\approx 1'' \approx
10^5 R_S$ and does not vary in time.  (2) X-ray 'flares' occurring at
a rate of $\approx 1$ day$^{-1}$ and lasting $\approx 10^3-10^4$ s.
The luminosity increases by a factor of few$-100$ during the flare and
the spectrum is quite hard ($\Gamma \approx 1.2$).  The flares are not
extended; in fact, the observed timescales argue that they arise close
to the BH, at $\lsim 10-100 R_S$.

\subsection{X-ray flares}

The X-ray flares are the most dramatic result from the {\it Chandra}
observations. Markoff et al. (2001) showed that the flares are
probably due to electron heating or acceleration, rather than a change
in the accretion rate onto the BH (otherwise there is too much
variation in other wavebands).  An obvious analogue is magnetic
reconnection in solar flares, which one could readily imagine
occurring in the inner part of the accretion flow close to the BH.
Given an injection of energy into the electrons, there are three
emission mechanisms that could, {\it a priori}, give rise to flares:
(1) bremsstrahlung, (2) synchrotron, and (3) inverse Compton.

Bremsstrahlung is attractive because it can naturally explain the very
hard spectrum of the flares. The problem is that, to produce a
luminosity of $L_{35} 10^{35}$ ergs s$^{-1}$ from a sphere of radius
$R$, the gas density must be $n \approx 10^9 L_{35}^{1/2}
T_{e,10}^{1/4} (R/10 R_S)^{-3/2}$ cm$^{-3}$, where $T_{e,10} =
T_e/10^{10} {\rm K}$.  For comparison, the ambient density in the
inner $10 R_S$ for a model with $\dot M \approx 10^{-8} \mpy$ is
$\approx 10^6$ cm$^{-3}$.  Equally importantly, bremsstrahlung
emission in RIAFs is dominated by very large radii $\sim R_{acc}$, not
small radii (e.g., Quataert \& Narayan 1999; see below).  Thus
bremsstrahlung appears unlikely to be responsible for the X-ray
flares.  Two possible ways out of this conclusion are (1) to posit
that the gas density increases by a factor of $\sim 10^3$ during the
flare, and does so preferentially in the very inner part of the
accretion flow (e.g., Liu \& Melia 2002).  It is unclear, however,
what would drive such large density changes, particularly since the
cooling time is so long that thermally driven instabilities are
unlikely to be important, (2) perhaps the accreting gas is a two-phase
medium, with a cooler, denser phase giving rise to the X-ray flares
(Yuan et al. 2003).

In contrast to bremsstrahlung, synchrotron emission can readily
account for the observed flares: if $\sim 10 \%$ of the electrons (by
energy) are accelerated into a power-law tail in the inner $\sim 10
R_S$, the radio-IR emission is essentially unchanged while electrons
with Lorentz factors $\gamma \sim 10^5$ can produce the X-ray emission
(e.g., Markoff et al. 2001).  Moreover, there is a 'natural'
explanation for the hard X-ray spectrum seen.  For RIAF models with
$\dot M \approx 10^{-8} \mpy$, the magnetic field strength close to
the BH is $\approx 20 B_{20}$ G.  The associated synchrotron cooling
time for electrons emitting in the {\it Chandra} band is $\approx 20
B_{20}^{-3/2}$ s.  Thus, unless $B \lsim 0.3$ G in the emitting
region, the cooling time is {less} than the duration of the flare and
there should be a cooling break in the electron distribution function
below the {\it Chandra} band.  For an injected distribution of
power-law electrons with $p_e < 2$, where $n(\gamma) \propto
\gamma^{-p_e}$, the distribution function for the population of cooled
electrons is $n(\gamma) \propto \gamma^{-2}$.  This implies a
synchrotron spectrum with $\Gamma = 1.5$, consistent with the typical
flare observed by {\it Chandra} (Baganoff et al. 2003b).

Synchrotron self-Compton emission can also explain the X-ray flares
observed by {\it Chandra}.  The dashed line in Figure \ref{fig:2}
shows a concrete example in which most of the electrons in a $\approx
3 R_S$ volume are accelerated into a power law distribution.  This
population of electrons produces synchrotron emission and also
upscatters synchrotron photons to produce a hard X-ray flare.  Note
that there is very little change to the radio or IR emission (the
dot-dashed line in Fig. \ref{fig:2} shows the baseline non-flaring
model in which power-law electrons have only $\approx 2 \%$ of the
electron thermal energy).


Finally, it is important to stress that in the models discussed here,
the duration of the flare is set by a dynamical or viscous timescale
in the inner $\sim 10 R_S$ of the accretion flow.  By contrast, there
is no explanation for the mean time between flares, $\approx 1$ day.
In addition, it is difficult to apply the ideas considered here to the
week-long, factor of few, sub-mm 'flares' observed by Tsuboi et
al. (1999) and Zhao et al. (2003).  In particular the duration of
these 'flares' is inconsistent with the characteristic timescales in
the sub-mm emitting region ($\lsim 10 R_S$).  One possibility is that
they are not directly related to the {\it Chandra} flares but are
instead due to small fluctuations in $\dot M$ set by dynamics in the
accretion flow at larger radii.
\subsection{Quiescent X-ray Emission}
The steady quiescent emission observed by {\it Chandra} is
qualitatively different from the flaring emission.  In particular, it
is softer ($\Gamma \approx 2.7$) and extended ($\approx 1'' \approx
10^5 R_S$).  The latter fact implies that it is a different emission
component since synchrotron and inverse Compton emission are produced
at small radii.

RIAF models naturally predict that the thermal bremsstrahlung emission
is dominated by large radii in the flow (e.g., Quataert \& Narayan
1999; Ozel \& Di Matteo 2000).  Given a density profile of the form
$\rho \propto R^{-3/2 + p}$ and a temperature profile $T \propto
R^{-1}$ (valid at large radii), the bremsstrahlung luminosity is
dominated by large radii: $L \propto R^3 \rho^2 T^{-1/2} \propto R^{2p
+ 1/2}$, assuming photon energies $\lsim kT(R)$.  The resulting
spectrum, adding up all radii, is $\Gamma \approx 3/2 + 2p$, i.e.,
$\nu L_\nu \propto \nu^{1/2 - 2p}$.

Thus a natural interpretation of the quiescent flux coincident with
Sgr A* is that it is bremsstrahlung from hot gas in the outer part of
the accretion flow that is resolved by {\it Chandra} (e.g., Yuan et
al. 2001; Quataert 2002).  This can account for the size of the
source, its lack of variability, and the possible presence of a
thermal X-ray line (Baganoff et al. 2003b).  The above expression for
the spectrum of the thermal emission would imply that $p \approx 1/2$
is required to explain the spectrum. There is, however, an important
problem with this straightforward interpretation: {\it Chandra}
spectra are extracted in a $\approx 1''$ region around Sgr A*.  This
is comparable to the Bondi accretion radius, $R_{acc}$ (\S1).  It is
thus incorrect to assume that observations of the extended emission
directly probe the ``accretion flow.''  Instead, they probe the
complex ``transition region'' between the accretion flow and the
ambient medium.

\begin{figure}[htb]
\includegraphics[width=.48\textwidth]{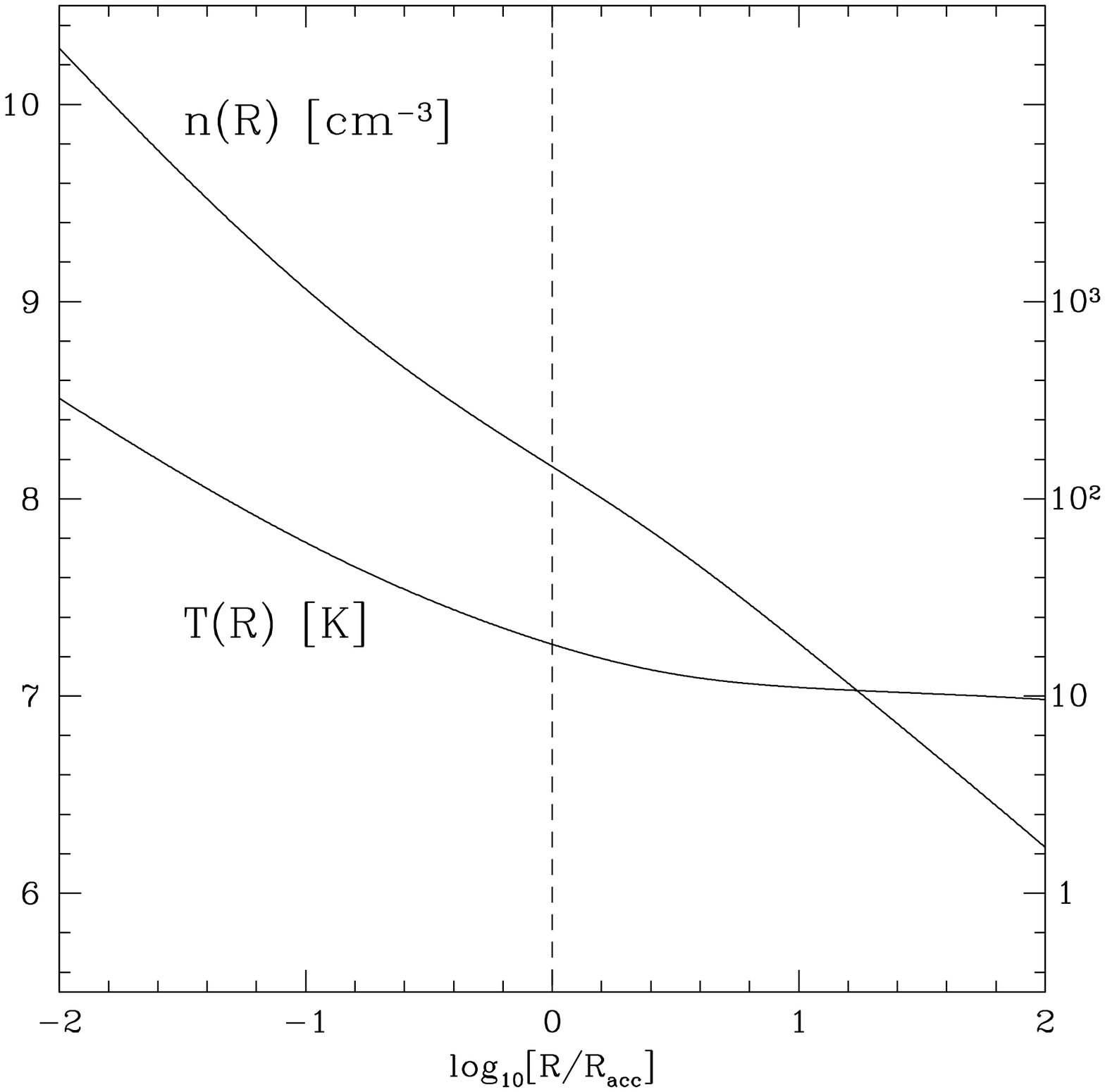}
\hfil
\includegraphics[width=.48\textwidth]{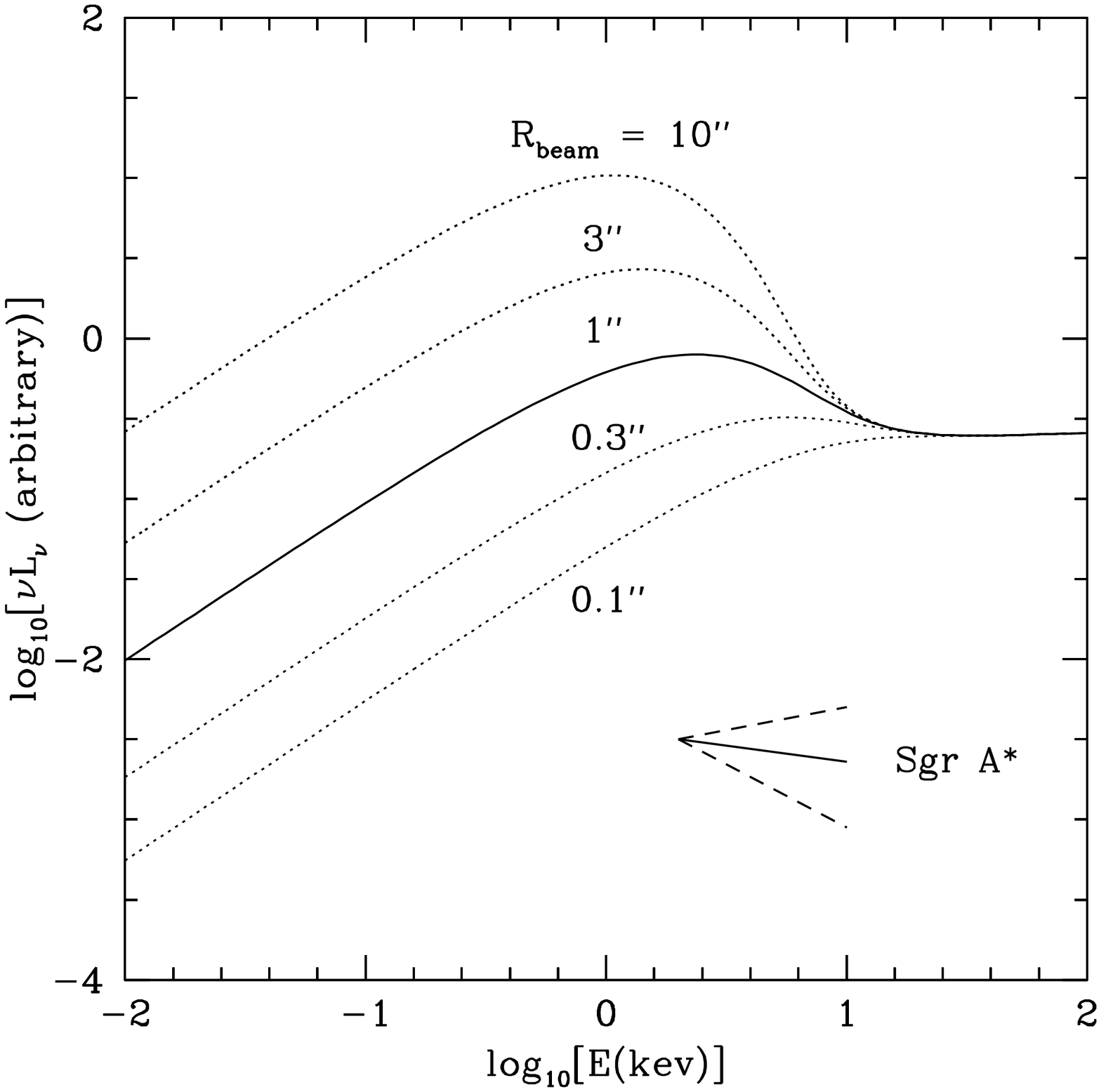}
\caption{{\it Left:} Observationally motivated models for $n(R)$ and
$T(R)$ for hot gas around Sgr A*; models are extrapolated to $R <
R_{acc} \approx 1''$ using Bondi accretion.  {\it Right:} X-ray
spectra based on the density and temperature profiles in the left
panel.  The $R_{beam} \approx 1''$ prediction is consistent with the
quiescent X-ray emission coincident with Sgr A*.}

\label{fig:3}
\end{figure}

Figure \ref{fig:3} illustrates the effects of a finite observing beam
relative to the Bondi accretion radius; the left panel shows a toy
model for the density and temperature as a function of radius around
Sgr A* (in units of $R_{acc} \approx 1''$).  The temperature is
measured to be $\approx 1$ keV at large radii.  The density profile on
$\gsim R_{acc}$ scales is adjusted to roughly reproduce the radial
variation of the observed diffuse X-ray emission (see Quataert 2002
for details).  For $R \lsim R_{acc}$ gas accretes as a Bondi flow,
with asymptotic ($R \ll R_{acc}$) scalings $\rho \propto R^{-3/2}$
(i.e., $p = 0$) and $T \propto R^{-1}$.\footnote{I chose a Bondi flow
for two reasons: (1) Bondi flows predict hard X-ray bremsstrahlung
spectra and are thus a good test case for assessing what effect the
{\it soft} X-ray emitting ambient medium around the BH has on
detecting the accretion component, (2) There are no good dynamical
models for how rotating RIAFs ``match'' onto an ambient medium outside
$R_{acc}$.}  Figure \ref{fig:3}b shows the X-ray spectra that would be
seen by a telescope with a beam-size $R_{beam}$.  For large beams,
$R_{beam} \approx 10''$, the spectrum is very soft and is dominated by
the ambient medium that has $T \approx 1$ keV.  For small beams,
$R_{beam} \approx 0.1''$, the spectrum is dominated by the accretion
flow and is quite hard, consistent with the above scalings.  For the
case applicable to Sgr A*, $R_{beam} \approx 1'' \approx R_{acc}$, the
emission is still relatively {\it soft}, consistent with the {\it
Chandra} observations.  This is in spite of the fact that the
underlying accretion flow produces a hard X-ray spectrum.

The upshot of Figure \ref{fig:3} is that the extended quiescent X-ray
source observed by {\it Chandra} appears broadly consistent with the
emission produced by gas on $R_{acc} \approx 1''$ scales.  This is gas
in the 'transition region' between the ambient medium and the
accretion flow, rather than the accretion flow itself.  Using these
results to constrain the dynamics of the accreting gas, e.g., the
radial density profile $p$, will require (1) better theoretical models
for the dynamics of the X-ray emitting gas on $1''$ scales, and (2)
tighter observational constraints, such as spectra as a function of
radius.

\section{Conclusions}

The Galactic Center represents a unique opportunity to probe the
dynamics of gas accreting onto a massive black hole, from the 'large'
scales on which gas is gravitationally captured by the BH to the
'small' scales close to the BH's horizon.  In this review, I have
tried to argue that radiatively inefficient accretion flow (RIAF)
models can provide a reasonably coherent picture of accretion onto Sgr
A* on all of these scales.

To summarize: hot gas in the central $\approx 1$ pc of the GC is
produced by the (shocked) winds from massive stars.  This gas is
gravitationally captured by the black hole on scales of $R_{acc}
\approx 1'' \approx 0.04$ pc.  The {\it Chandra} detection of an
extended soft X-ray component coincident with Sgr A* may be direct
evidence for this gravitationally captured gas (\S3.2).  The net rate
at which gas accretes through the BH's horizon is, however, much less
than the canonical Bondi estimate for the rate at which gas is
gravitationally captured by the BH ($\dot M_B \approx 10^{-5} \mpy$).
The Bondi estimate neglects the angular momentum of the accreting gas,
which is likely to be a very poor assumption.  Numerical simulations
(and analytical models) of {\it rotating} RIAFs find that $\dot M \ll
\dot M_B$ (\S2).  This conclusion is theoretically attractive because
it implies that the radiative efficiency of Sgr A* need not be as low
as $\sim 10^{-6}$ as in Bondi and ADAF models (which has always been
difficult to reconcile with the expectation that electron heating and
acceleration would be important in the collisionless magnetized plasma
close to the BH).  A low accretion rate is also strongly suggested by
the detection of linear polarization in the sub-mm emission from Sgr
A*: Faraday rotation constrains the gas density and magnetic field
strength close to the BH and argues for $\dot M \sim 10^{-8} \mpy \ll
\dot M_B$, in good agreement with the inference from RIAF models.
Finally, RIAF models with $\dot M \sim 10^{-8} \mpy$ can explain the
basic spectral properties of Sgr A* (see, e.g., Figs. \ref{fig:1} \&
\ref{fig:2}).  In particular, the X-ray flares seen by {\it Chandra}
may be due to synchrotron or Inverse-Compton emission produced by
relativistic electrons accelerated in the inner $\sim 10 R_S$ of the
accretion flow (\S3.1).

It is important to stress that, although $\dot M \ll \dot M_B$ is both
theoretically and observationally favored, all of the models
considered here are still ``radiatively inefficient,'' and have
efficiencies much less than the canonical thin disk value of $10 \%$;
e.g., for $\dot M \sim 10^{-8} \mpy$, $L \sim 10^{-3} \dot M c^2$.  In
fact, all of the physics highlighted in \S2 that suppresses the
accretion rate with respect to the Bondi estimate requires a
relatively low efficiency and would not operate in a thin disk.

There are several important issues that I have not addressed in this
review.  To cite two that clearly require further study: (1) are the
{\it Chandra} flares due to, e.g., reconnection or turbulence in the
accretion flow, or are they telling us something more fundamental
about the dynamics close to the BH? (2) Both jet and RIAF models can
explain the basic spectral properties of Sgr A*.  How can we
distinguish between these two components, both of which are almost
certainly present? E.g., is the linear polarization detection, which
requires a relatively coherent magnetic field, consistent with the
magnetic field seen in RIAF simulations?  Or does it instead require
an additional 'jet' component?

\begin{acknowledgement}
I thank Ramesh Narayan and Feng Yuan for useful discussions.  This
work is supported by NASA Grant NAG5-12043, NSF Grant AST-0206006, and
an Alfred P. Sloan Foundation Fellowship.
\end{acknowledgement}


\begin{thebibliography}{10}
\bibitem{} Aitken, D. K. et al., 2000, ApJ, 534, L173
\bibitem{} Agol, E., 2000, ApJ, 538, L121
\bibitem{} Baganoff, F. et al., 2001, Nature, 413, 45
\bibitem{} Baganoff, F. et al., 2003a, ApJ, in press
\bibitem{} Baganoff, F. et al., 2003b, these proceedings
\bibitem{} Beckert, T. \& Falcke, H., 2002, A\&A, 388, 1106
\bibitem{} Beckert, T. \& Falcke, H., 2003, these proceedings
\bibitem{} Blandford, R.D. \& Begelman, M.C., 1999, MNRAS, 303, L1
\bibitem{} Bondi, H., 1952, MNRAS, 112, 195
\bibitem{} Bower, G., Wright, M. C. H., Falcke, H., \& Backer, D., 2003, ApJ, in press (astro-ph/0302227)
\bibitem{} Falcke, H \& Markoff, S., 2000, A\&A, 362, 113
\bibitem{} Ghez, A. et al. 2003, these proceedings
\bibitem{} Hawley, J. F. \& Balbus, S. A., 2002, ApJ, 573, 738
\bibitem{} Herrnstein, R. M. \&  Ho, P. T., 2002, ApJ, 579, L83
\bibitem{} Ichimaru, S. 1977, ApJ, 214, 840
\bibitem{} Igumenshchev, I. V., \& Abramowicz, M. A. 1999, MNRAS, 303, 309
\bibitem{} Igumenshchev, I. V., \& Abramowicz, M. A. 2000, ApJS, 130, 463
\bibitem{} Igumenshchev, I. V., Abramowicz, M. A., \& Narayan, R. 2000, ApJ, 537, L27
\bibitem{} Igumenshchev, I. V., Abramowicz, M. A., \& Narayan, R. 2003,
ApJ submitted (astro-ph/0301402)
\bibitem{} Koratkar, A. \& Blaes, O. 1999, PASP, 111, 1
\bibitem{} Krabbe, A., Genzel, R., Drapatz, S., \& Rotaciuc, V., 1991, ApJ, 382, L19
\bibitem{} Liu, S. \& Melia, F., 2002, ApJ, 566, L77
\bibitem{} Mahadevan, R., 1998, Nature, 394, 651
\bibitem{} Manmoto, T., Mineshige, S., \& Kusunose, M., 1997, ApJ, 489, 791
\bibitem{} Markoff, S., Falcke, H., Yuan, F., \& Biermann, P. L., 2001, A\&A, 379, L13
\bibitem{} Melia, F., 1992, ApJ, 387, L25
\bibitem{} Narayan, R. 2002, in {\it Lighthouses of the Universe},
ed. M. Gilfanov, \& R. Sunyaev (Heidelberg: Springer-Verlag), in press,
astro-ph/0201260
\bibitem{} Narayan, R., Igumenshchev, I. V., \& Abramowicz, M. A.
     2000, ApJ, 539, 798
\bibitem{} Narayan, R., Quataert, E., Igumenshchev, I. V., \& Abramowicz, M. A. 2002, ApJ, 577, 295
\bibitem{} Narayan, R. et al., 1998, ApJ, 492, 554
\bibitem{} Narayan, R., Yi, I., \& Mahadevan, R., 1995, Nature, 374, 623
\bibitem{} Narayan, R., \& Yi, I. 1994, ApJ, 428, L13
\bibitem{} Najarro, F. et al., 1997, A\&A, 325, 700
\bibitem{} Nayakshin, S., 2003, these proceedings
\bibitem{} Ozel, F. \& Di Matteo, T., 2001, ApJ, 563, 276
\bibitem{} Ozel, F., Psaltis, D., \& Narayan, R., 2000, ApJ, 541, 234
\bibitem{} Quataert, E., 2002, ApJ, 575, 855
\bibitem{} Quataert, E., Dorland, W., \& Hammett, G., 2002, ApJ, 577, 524
\bibitem{} Quataert, E. \& Gruzinov, A., 1999, ApJ, 520, 248
\bibitem{} Quataert, E. \& Gruzinov, A., 2000a, ApJ, 539, 809
\bibitem{} Quataert, E. \& Gruzinov, A., 2000b, ApJ, 545, 842 
\bibitem{} Quataert, E. \& Narayan, R., 1999, ApJ, 520, 298 
\bibitem{} Rees, M. J., Begelman, M. C., Blandford, R. D., \& Phinney, E. S., 1982, Nature, 295, 17 
\bibitem{} Ruszkowski, M. \& Begelman, M. C., 2002, ApJ, 573, 485
\bibitem{} Sch\"odel, R. et al., 2002, Nature, 419, 694
\bibitem{} Shakura, N. I., \& Sunyaev, R. A., 1973, A\&A, 24, 337 
\bibitem{} Stone, J. M., Pringle, J. E., \& Begelman, M. C. 1999, MNRAS,
310, 1002
\bibitem{} Stone, J. \& Pringle, J. E., 2001, MNRAS, 322, 461
\bibitem{} Tsuboi, M., Miyazaki, A., \& Tsutsumi, T., 1999, in {\it
The Central Parsecs of the Galaxy, ASP Conference Series}, ed. Heino
Falcke et al., Vol, 186, p. 105
\bibitem{} Yuan, F., Markoff, S., \& Falcke, H., 2002, A\&A, 383, 854
\bibitem{} Yuan, F., Quataert, E., \& Narayan, R., 2003, in prep.
\bibitem{} Zhao, J. et al., 2003, ApJ Letters in press (astro-ph/0302062)
\end{thebibliography}
\end{document}